\documentclass{cicirello}
\pdfoutput=1

\usepackage[numbers]{natbib}
\usepackage{doi}

\title{On the Average Runtime of an Open Source Binomial Random Variate Generation Algorithm}
\author{Vincent A. Cicirello}
\orcid{0000-0003-1072-8559}
\reportnum{ALG-24-007}
\reporturl{https://reports.cicirello.org/24/007/ALG-24-007.pdf}
\copyrightyear{2024}
\date{March 2024}

\address{Computer Science\\
Stockton University\\
Galloway, NJ 08205 USA\\
\url{https://www.cicirello.org/}
}

\ACM{F.2.1; G.3; G.4; I.1.2}

\abstract{The BTPE algorithm (Binomial, Triangle, Parallelogram, Exponential)
of Kachitvichyanukul and Schmeiser is one of the faster and more widely utilized
algorithms for generating binomial random variates. Cicirello's open source
Java library, $\rho\mu$, includes an implementation of BTPE as well as a variety
of other random number related utilities. In this report, I explore the average
case runtime of the BTPE algorithm when generating random values from binomial
distribution $B(n,p)$. Beginning with Kachitvichyanukul and Schmeiser's
formula for the expected number of acceptance-rejection sampling iterations,
I analyze the limit behavior as $n$ approaches infinity, and show
that the average runtime of BTPE converges to a constant. I instrument 
the open source Java implementation from the $\rho\mu$ library to experimentally 
validate the analysis.}

\begin{document}

\maketitle

\section{Introduction}

In this report, I explore the average runtime behavior of binomial random variate 
generation within the open source Java library $\rho\mu$~\cite{cicirello2022joss}.
The $\rho\mu$ library provides a variety of random number related enhancements
beyond what is provided by the Java API itself. The core functionality of $\rho\mu$ is 
provided through a hierarchy of wrapper classes. This set of $\rho\mu$'s wrapper classes
directly corresponds to the hierarchy of random number generator 
interfaces introduced in Java 17~\cite{Steele2017}. In some cases, $\rho\mu$ overrides
Java's random number generators with faster algorithms, such as for random integers 
subject to a bound. In other cases, $\rho\mu$ adds functionality, such as additional 
distributions, i.e., the binomial, Cauchy, among others~\cite{cicirello2022joss}. 
The initial motivation of the $\rho\mu$ library was to provide a source of enhanced and 
more efficient randomness for other libraries, such as JavaPermutationTools~\cite{cicirello2018joss} 
and Chips-n-Salsa~\cite{cicirello2020joss}.

Among the randomness enhancements of the $\rho\mu$ library is support for generating
binomial random variates, which are important to many 
applications~\cite{Singh2022,Shah2022,Garcia2022,Zhang2020,Naimi2020,Arshad2019,Khan2019,Wang2019,Kuhl2017}.
There are many algorithms for generating binomial random 
variates~\cite{Kachitvichyanukul1988,Kachitvichyanukul1989,Relles1972,Ahrens1974,Kuhl2017,Knuth3}.
The $\rho\mu$~\cite{cicirello2022joss} library implements the BTPE Algorithm (Binomial, Triangle, 
Parallelogram, Exponential)~\cite{Kachitvichyanukul1988}, and falls back upon the inverse 
transform~\cite{Kachitvichyanukul1988,Knuth3} for cases that cannot be handled by BTPE.

The runtime of many algorithms for generating random values from binomial distribution $B(n,p)$
grows with some function of $n$ and $p$. For example, the runtime of the inverse transform 
method is $O(np)$~\cite{Kachitvichyanukul1988,Knuth3}. One technique many binomial random variate
algorithms utilize is acceptance--rejection sampling~\cite{Flury1990}, and the iterations of
such sampling is often what leads to such runtimes. BTPE does use acceptance--rejection sampling,
but seems to require relatively few iterations. At the time of introduction,
Kachitvichyanukul and Schmeiser determined the expected number of iterations as a function of
$n$ and $p$~\cite{Kachitvichyanukul1988} (see Section~\ref{sec:BTPEprelim}). However, the
aim of my present report is to provide simpler insight into the average case runtime than 
is provided by Kachitvichyanukul and Schmeiser's formula.

In this report, I explore the limit of Kachitvichyanukul and Schmeiser's formula
as $n$ approaches infinity (Section~\ref{sec:limit}), for two cases of $p$, including
the minimum $p$ supported by BTPE (i.e., $p=\frac{10}{n}$), which maximizes the
expected number of rejection sampling iterations, and $p=0.5$, which is when the 
expected number of rejection sampling iterations is minimized. Both cases lead to 
average runtimes that are constants in the limit for large $n$. In 
Section~\ref{sec:experiments}, I validate the findings experimentally using
the $\rho\mu$ library, and utilizing my prior empirical study~\cite{cicirello2023asec}.
I discuss the findings and offer conclusions in Section~\ref{sec:conclusions},
including the main result that the average runtime of BTPE is $\Theta(1)$.
I now begin in Section~\ref{sec:prelim} with some required background.

\section{Preliminaries}\label{sec:prelim}

\subsection{Rejection Sampling Iterations of BTPE}\label{sec:BTPEprelim}

The BTPE~\cite{Kachitvichyanukul1988} algorithm separates the binomial distribution
$B(n,p)$ into four parts, using triangular functions in the middle portions, and 
exponential functions in the tails, and it uses acceptance--rejection 
sampling~\cite{Flury1990}. For complete details of BTPE, which are beyond the 
scope of this paper, I refer the reader to the article that introduced 
it~\cite{Kachitvichyanukul1988}. 

To generate a random value from binomial distribution $B(n,p)$, each acceptance--rejection 
iteration of BTPE generates two random values from $U(0,1)$, i.e., uniformly
distributed over the interval $[0.0,1.0)$. When they introduced BTPE, Kachitvichyanukul
and Schmeiser determined that the expected number of iterations, $E[I]$ of BTPE 
is~\cite{Kachitvichyanukul1988}:
\begin{equation}\label{eq:iter}
E[I] = p_4 \binom{n}{M} r^{M} (1-r)^{n-M}  ,
\end{equation}
and since each iteration generates two random uniform values from $U(0,1)$, the
expected number of uniform variates, $E[V]$, required by BTPE is thus:
\begin{equation}\label{eq:uniform}
E[V] = 2 p_4 \binom{n}{M} r^{M} (1-r)^{n-M}.
\end{equation}

Equations~\ref{eq:iter} and~\ref{eq:uniform} involve $r$, $M$, and $p_4$, which in turn
depend upon a few others, all of which are functions of $n$ and $p$. Kachitvichyanukul 
and Schmeiser~\cite{Kachitvichyanukul1988} define these as follows:
\begin{gather}
r = \min(p, 1-p) ,\\
q = 1 - r ,\\
f_M = nr + r ,\\
M = \lfloor f_M \rfloor ,\\
p_1 = \left\lfloor 2.195 \sqrt{nrq} - 4.6 q \right\rfloor + 0.5 ,\\
x_M = M + 0.5 ,\\
x_L = x_M - p_1 ,\\
x_R = x_M + p_1 ,\\
c = 0.134 + \frac{20.5}{15.3 + M} ,\\
a_L = \frac{f_M - x_L}{f_M - x_L r} ,\\
\lambda_L = a_L \left( 1 + \frac{a_L}{2} \right) ,\\
a_R = \frac{x_R - f_M}{x_R q} ,\\
\lambda_R = a_R \left( 1 + \frac{a_R}{2} \right) ,\\
p_2 = p_1 (1 + 2c) ,\\
p_3 = p_2 + \frac{c}{\lambda_L} ,\\
p_4 = p_3 + \frac{c}{\lambda_R} .
\end{gather}

Also note that BTPE is only relevant when: $\min(p, 1-p) \geq \frac{10}{n}$. 
Kachitvichyanukul and Schmeiser recommend using the inverse transform 
method~\cite{Knuth3} as a fallback for cases when BTPE is not relevant. This is
what is done in the implementation in the $\rho\mu$ library~\cite{cicirello2022joss}.

The expected number of acceptance--rejection sampling iterations, expressed in 
Equation~\ref{eq:uniform}, is maximized when $\min(p, 1-p)$ is minimal.
Thus, the expected number of rejection sampling iterations is maximized for the 
cases $p=\frac{10}{n}$ and $p=\frac{n-10}{n}$. Due to symmetry, these two cases 
lead to the same expected number of rejection sampling iterations. The nearer $p$ 
is to $0.5$, the fewer iterations BTPE requires on average.
Table~\ref{tab:predicted} shows how the expected number of required uniform
variates changes as $n$ increases for these two cases of $p$. It includes
the minimum $n$ supported by BTPE (e.g., $n=20$), and then considers $n$ at
increasing powers of two to show how the number of uniform variates used by BTPE
to generate one binomial random variate varies as $n$ increases rapidly.

\begin{table}[t]
\centering
\caption{Expected number of uniform variates to generate one binomial random variate
as predicted by Equation~\ref{eq:uniform} as $n$ increases.}\label{tab:predicted}
\begin{tabular}{c|cc}\hline
	&	\multicolumn{2}{c}{$E[V]$ from Equation~\ref{eq:uniform}} \\
$n$			& $p\in\left\{\frac{10}{n},\frac{n-10}{n}\right\}$	& $p=0.5$ \\ \hline
$20$ & 3.996 & 3.996 \\
$2^{5}$ & 3.837 & 3.599 \\
$2^{6}$ & 3.792 & 3.337 \\
$2^{7}$ & 3.790 & 2.985 \\
$2^{8}$ & 3.793 & 2.632 \\
$2^{9}$ & 3.796 & 2.420 \\
$2^{10}$ & 3.797 & 2.317 \\
$2^{11}$ & 3.798 & 2.307 \\
$2^{12}$ & 3.798 & 2.276 \\
$2^{13}$ & 3.799 & 2.300 \\
$2^{14}$ & 3.799 & 2.299 \\
$2^{15}$ & 3.799 & 2.300 \\
$2^{16}$ & 3.799 & 2.302 \\
$2^{17}$ & 3.799 & 2.310 \\
$2^{18}$ & 3.799 & 2.310 \\
$2^{19}$ & 3.799 & 2.313 \\
$2^{20}$ & 3.799 & 2.314 \\\hline
\end{tabular}
\end{table}

\subsection{Stirling's Formula}

While computing the limiting behavior of BTPE, we will encounter some
factorials. Stirling's formula~\cite{CLRS} will be useful in the analysis. Stirling's
formula is as follows:
\begin{equation}\label{eq:factorial}
n! = \sqrt{2 \pi n} \left(\frac{n}{e}\right)^n e^{\alpha_n} ,
\end{equation}
where
\begin{equation}\label{eq:alpha}
\frac{1}{12 n + 1} < \alpha_n < \frac{1}{12 n} . 
\end{equation}

A consequence of using Stirling's formula is that it will
introduce terms involving $e^{\alpha_n}$, either generally for $n$
or in other cases for specific values of $n$. The following observations
will be useful:
\begin{gather}
\lim_{n \to \infty} \alpha_n = 0 \: , \\
\lim_{n \to \infty} \alpha_{n/2} = 0 \: , \\
\lim_{n \to \infty} \alpha_{n - 10} = 0 \: , \\
\lim_{n \to \infty} e^{\alpha_n} = 1 \: , \\
\lim_{n \to \infty} e^{\alpha_{n/2}} = 1 \: , \\
\lim_{n \to \infty} e^{\alpha_{n - 10}} = 1 \: .
\end{gather}

\section{Limit Analysis}\label{sec:limit}

Let's now proceed to compute the limit:
\begin{equation}\label{eq:limit}
\lim_{n \to \infty} E[V] = \lim_{n \to \infty} 2 p_4 \binom{n}{M} r^{M} (1-r)^{n-M} ,
\end{equation}
for two cases, when $p=\frac{10}{n}$ and when $p=0.5$ in Sections~\ref{sec:maxiter}
and~\ref{sec:miniter}, respectively.

\subsection{Case 1: Maximum Rejection Sampling Iterations}\label{sec:maxiter}

Consider the limit from Equation~\ref{eq:limit} when $p=\frac{10}{n}$,
which is the lowest $p$ (in terms of $n$) supported by BTPE.
Begin by computing the values of the various constants from Section~\ref{sec:BTPEprelim}
in terms of $n$ and $p=\frac{10}{n}$ as follows:
\begin{gather}
r = \min(p, 1-p) = \frac{10}{n} \:,\\
q = 1 - r = \frac{n - 10}{n} \:,\\
f_M = nr + r = n \left(\frac{10}{n}\right) + \frac{10}{n} = 10 + \frac{10}{n} \:,\\
M = \lfloor f_M \rfloor = \left\lfloor 10 + \frac{10}{n} \right\rfloor = 10 \:,\\
c = 0.134 + \frac{20.5}{15.3 + M} = 0.134 + \frac{20.5}{15.3 + 10} \approx 0.944 \:,\\
x_M = M + 0.5 = 10 + 0.5 = 10.5 \:.
\end{gather}

From Equation~\eqref{eq:alpha}, find the following, which we need later:
\begin{gather}
\frac{1}{121} < \alpha_{10} < \frac{1}{120} \: ,\\
\alpha_{10} \approx 0.0083 \: ,\\
e^{\alpha_{10}} \approx 1.008 \: .
\end{gather}

We will also need the following limit:
\begin{equation}
\begin{split}
\lim_{n \to \infty} p_1 & = \lim_{n \to \infty} \left\lfloor 2.195 \sqrt{nrq} - 4.6 q \right\rfloor + 0.5 \\
	& = \lim_{n \to \infty} \left\lfloor 2.195 \sqrt{n \left(\frac{10}{n}\right) \left(\frac{n-10}{n}\right)} - 4.6 \left(\frac{n-10}{n}\right) \right\rfloor + 0.5 \\
	& = \lim_{n \to \infty} \left\lfloor 2.195 \sqrt{10 \left(\frac{n-10}{n}\right)} - 4.6 \left(\frac{n-10}{n}\right) \right\rfloor + 0.5 \\
	& = \left\lfloor 2.195 \sqrt{10} - 4.6 \right\rfloor + 0.5 \\
	& = 2.5 \: .
\end{split}
\end{equation}

We will also need the limits of $\lambda_L$ and $\lambda_R$ as $n$ approaches infinity,
as follows:
\begin{equation}
\begin{split}
\lim_{n \to \infty} \lambda_L & = \lim_{n \to \infty} a_L \left( 1 + \frac{a_L}{2} \right) \\
	& = \lim_{n \to \infty} \left( \frac{f_M - x_L}{f_M - x_L r} \right) \left( 1 + \frac{f_M - x_L}{2 (f_M - x_L r)} \right) \\
	& = \lim_{n \to \infty} \left( \frac{f_M - x_M + p_1}{f_M - r (x_M - p_1) } \right) \left( 1 + \frac{f_M - x_M + p_1}{2 (f_M - r(x_M - p_1))} \right) \\
	& = \lim_{n \to \infty} \left( \frac{f_M - 8}{f_M - 8 r} \right) \left( 1 + \frac{f_M - 8}{2 (f_M - 8 r)} \right) \\
	& = \lim_{n \to \infty} \left( \frac{\frac{10n + 10}{n} - 8}{\frac{10n + 10}{n} - \frac{80}{n}} \right) \left( 1 + \frac{\frac{10n + 10}{n} - 8}{2 (\frac{10n + 10}{n} - \frac{80}{n})} \right) \\
	& = \left( \frac{10-8}{10} \right) \left( 1 + \frac{10-8}{20} \right) \\
	& = 0.22
\end{split}
\end{equation}
and
\begin{equation}
\begin{split}
\lim_{n \to \infty} \lambda_R & = \lim_{n \to \infty} a_R \left( 1 + \frac{a_R}{2} \right) \\
	& = \lim_{n \to \infty} \left( \frac{x_R - f_M}{x_R q} \right) \left( 1 + \frac{x_R - f_M}{2 x_R q} \right) \\
	& = \lim_{n \to \infty} \left( \frac{x_M + p_1 - f_M}{q(x_M + p_1)} \right) \left( 1 + \frac{x_M + p_1 - f_M}{2 q(x_M + p_1)} \right) \\
	& = \lim_{n \to \infty} \left( \frac{13 - f_M}{13q} \right) \left( 1 + \frac{13 - f_M}{26q} \right) \\
	& = \lim_{n \to \infty} \left( \frac{13 - \frac{10n + 10}{n}}{\frac{13n-130}{n}} \right) \left( 1 + \frac{13 - \frac{10n + 10}{n}}{\frac{26n-260}{n}} \right) \\
	& = \left( \frac{13 - 10}{13} \right) \left(1 + \frac{13 - 10}{26} \right) \\
	& \approx 0.257 .
\end{split} 
\end{equation}

We finally can compute the limit of Equation~\eqref{eq:limit}:
\begin{equation}\label{eq:limit2}
\begin{split}
\lim_{n \to \infty} E[V] & = \lim_{n \to \infty} 2 p_4 \binom{n}{M} r^{M} (1-r)^{n-M} \\
	& = \lim_{n \to \infty} 2 p_4 \binom{n}{10} \left(\frac{10}{n}\right)^{10} \left(\frac{n - 10}{n}\right)^{n-10} \\
	& = \lim_{n \to \infty} 2 p_4 \left( \frac{n!}{10! (n-10)!} \right) \left(\frac{10}{n}\right)^{10} \left(\frac{n - 10}{n}\right)^{n-10} \\
	& = \lim_{n \to \infty} 2 p_4 \left( \frac{\sqrt{\frac{n}{(n-10)}} \, n^n \, e^{\alpha_n}}{\sqrt{20 \pi} \, 10^{10} (n-10)^{n-10} \, e^{\alpha_{n-10}} \, e^{\alpha_{10}}} \right) \left(\frac{10}{n}\right)^{10} \left(\frac{n - 10}{n}\right)^{n-10} \\
	& = \lim_{n \to \infty} 2 p_4 \left( \frac{\sqrt{\frac{n}{(n-10)}}}{\sqrt{20 \pi} \, e^{\alpha_{10}}} \right)  \approx \lim_{n \to \infty} 2 p_4 \left( \frac{\sqrt{\frac{n}{(n-10)}}}{1.008 \sqrt{20 \pi}} \right) \approx \lim_{n \to \infty} 0.2503 p_4 \sqrt{\frac{n}{(n-10)}} \\
	& \approx \lim_{n \to \infty} 0.2503 p_4 \\
	& \approx \lim_{n \to \infty} 0.2503 \left( p_1 (1 + 2c) + \frac{c}{\lambda_L} + \frac{c}{\lambda_R} \right) \\
	& \approx \lim_{n \to \infty} 0.2503 \left( 2.888 \, p_1 + \frac{0.944}{\lambda_L} + \frac{0.944}{\lambda_R} \right) \\
	& \approx \lim_{n \to \infty} 0.2503 \left( 7.22 + \frac{0.944}{\lambda_L} + \frac{0.944}{\lambda_R} \right) \\
	& \approx 0.2503 \left( 7.22 + \frac{0.944}{0.22} + \frac{0.944}{0.257} \right) \approx 0.2503 \left( 7.22 + 4.291 + 3.673 \right) \\
	& \approx 3.801 .
\end{split}
\end{equation}

Thus, the expected number of uniform random variates required by BTPE to generate
one binomial random variate converges to approximately 3.801 as $n$ grows large (approximately
1.90 rejection sampling iterations on average) for the case of minimum supported $p$. 
Observe in Table~\ref{tab:predicted} that BTPE approaches this limit case rapidly.

\subsection{Case 2: Central Case}\label{sec:miniter}

Now consider the limit from Equation~\ref{eq:limit} for the case of $p=0.5$.

Let us begin by computing some of the constants from Section~\ref{sec:BTPEprelim}
for this case. When computing $M$, assume that $n$ is even, without loss
of generality:
\begin{gather}
r = \min(p, 1-p) = \frac{1}{2} ,\\
q = 1 - r = \frac{1}{2} ,\\
f_M = nr + r = \frac{n+1}{2} ,\\
M = \lfloor f_M \rfloor = \frac{n}{2} \:.
\end{gather}

Now substitute these, and other Section~\ref{sec:BTPEprelim} definitions, into 
Equation~\ref{eq:limit} and simplify:
\begin{equation}\label{eq:p5limit1}
\begin{split}
\lim_{n \to \infty} E[V] & = \lim_{n \to \infty} 2 p_4 \binom{n}{M} r^{M} (1-r)^{n-M} \\
	& = \lim_{n \to \infty} 2 p_4 \binom{n}{\frac{n}{2}} \left(\frac{1}{2}\right)^{\frac{n}{2}} \left(1-\frac{1}{2}\right)^{n-\frac{n}{2}} \\
	& = \lim_{n \to \infty} 2 p_4 \left(\frac{n!}{\left(\frac{n}{2}\right)!\left(\frac{n}{2}\right)!}\right) \left(\frac{1}{2}\right)^{n} \\
	& = \lim_{n \to \infty} 2 p_4 \left(\frac{\sqrt{2 \pi n}\left(\frac{n}{e}\right)^{n} e^{\alpha_n}}{\sqrt{\pi n}\left(\frac{n}{2e}\right)^{n/2} e^{\alpha_{n/2}} \sqrt{\pi n}\left(\frac{n}{2e}\right)^{n/2} e^{\alpha_{n/2}}}\right) \left(\frac{1}{2}\right)^{n} \\
	& = \lim_{n \to \infty} 2 p_4 \left(\frac{\sqrt{2 \pi n}\left(\frac{n}{e}\right)^{n} e^{\alpha_n}}{\pi n \left(\frac{n}{e}\right)^{n} \left(\frac{1}{2}\right)^{n} e^{2\alpha_{n/2}}}\right) \left(\frac{1}{2}\right)^{n} \\
	& = \lim_{n \to \infty} 2 p_4 \left(\frac{\sqrt{2/\pi}}{\sqrt{n}}\right) \left(\frac{e^{\alpha_n}}{e^{2\alpha_{n/2}}} \right) \\
	& \approx \lim_{n \to \infty} 1.596 \left(\frac{p_4}{\sqrt{n}}\right) \\
	& \approx \lim_{n \to \infty} 1.596 \left(\frac{1}{\sqrt{n}}\right) \left(p_1(1+2c) + \frac{c}{\lambda_L} + \frac{c}{\lambda_R} \right) \\
	& \approx \lim_{n \to \infty} 1.596 \left(\frac{1}{\sqrt{n}}\right) \left(p_1(1+2c) + \frac{2c}{\lambda_R} \right) \\
	& \approx \lim_{n \to \infty} 1.596 \left(\frac{p_1 (1+2c)}{\sqrt{n}} + \frac{2c}{\sqrt{n}\lambda_R} \right) .
\end{split}
\end{equation}
The next-to-last step above utilizes the fact that $\lambda_R = \lambda_L$ when $p=0.5$.

Before continuing to simplify Equation~\ref{eq:p5limit1}, compute the following limits. First:
\begin{equation}\label{eq:c5limit}
\lim_{n \to \infty} c = \lim_{n \to \infty} 0.134 + \frac{20.5}{15.3 + M} = \lim_{n \to \infty} 0.134 + \frac{20.5}{15.3 + \frac{n}{2}} = 0.134 .
\end{equation}
Now, compute the limit $\lim_{n \to \infty} \frac{p_1}{\sqrt{n}}$. Since $p_1$ involves a floor, 
compute bounds on this limit. Although we will see that the upper and lower bounds are equal, and thus
via the squeeze theorem is also equal to our target limit.
\begin{equation}\label{eq:p1p5limit}
\begin{split}
\lim_{n \to \infty} \frac{\left\lfloor 2.195 \sqrt{nrq} - 4.6 q \right\rfloor + 0.5}{\sqrt{n}} \: & \leq \lim_{n \to \infty} \frac{p_1}{\sqrt{n}} \leq \: \lim_{n \to \infty} \frac{\left\lfloor 2.195 \sqrt{nrq} - 4.6 q \right\rfloor + 0.5}{\sqrt{n}} \\
\lim_{n \to \infty} \frac{\left\lfloor 2.195 \sqrt{n/4} - 2.3 \right\rfloor + 0.5}{\sqrt{n}} \: & \leq \lim_{n \to \infty} \frac{p_1}{\sqrt{n}} \leq \: \lim_{n \to \infty} \frac{\left\lfloor 2.195 \sqrt{n/4} - 2.3 \right\rfloor + 0.5}{\sqrt{n}} \\
\lim_{n \to \infty} \frac{\left\lfloor 1.0975 \sqrt{n} - 2.3 \right\rfloor + 0.5}{\sqrt{n}} \: & \leq \lim_{n \to \infty} \frac{p_1}{\sqrt{n}} \leq \: \lim_{n \to \infty} \frac{\left\lfloor 1.0975 \sqrt{n} - 2.3 \right\rfloor + 0.5}{\sqrt{n}} \\
\lim_{n \to \infty} \frac{1.0975 \sqrt{n} - 2.8}{\sqrt{n}} \: & \leq \lim_{n \to \infty} \frac{p_1}{\sqrt{n}} \leq \: \lim_{n \to \infty} \frac{1.0975 \sqrt{n} - 1.8}{\sqrt{n}} \\
1.0975 \: & \leq \lim_{n \to \infty} \frac{p_1}{\sqrt{n}} \leq \: 1.0975
\end{split}
\end{equation}
The next limit that we need is $\lim_{n \to \infty} \frac{2}{\sqrt{n}\lambda_R}$ as follows:
\begin{equation}\label{eq:lambdap5limit}
\begin{split}
\lim_{n \to \infty} \frac{2}{\sqrt{n}\lambda_R} & = \lim_{n \to \infty} \frac{2}{\sqrt{n} \left( a_R \left( 1 + \frac{a_R}{2} \right)\right)} \\
	& = \lim_{n \to \infty} \frac{2 x_R q}{\sqrt{n} (x_R - f_M) \left( 1 + \frac{x_R - f_M}{2 x_R q} \right)} \\
	& = \lim_{n \to \infty} \frac{x_R}{\sqrt{n} (x_R - f_M) \left( 1 + \frac{x_R - f_M}{x_R} \right)}
\end{split}
\end{equation}
Now consider the following:
\begin{equation}
x_R - f_M = x_M + p_1 - \frac{n+1}{2} = M + \frac{1}{2} + p_1 - \frac{n+1}{2} = \frac{n}{2} + \frac{1}{2} + p_1 - \frac{n+1}{2} = p_1 .
\end{equation}
Now continue simplifying Equation~\ref{eq:lambdap5limit}, keeping Equation~\ref{eq:p1p5limit} in mind 
as well as that $p_1 \in \Theta(\sqrt{n})$ as follows:
\begin{equation}\label{eq:lambdap5limit2}
\begin{split}
\lim_{n \to \infty} \frac{2}{\sqrt{n}\lambda_R} & = \lim_{n \to \infty} \frac{x_R}{\sqrt{n} (x_R - f_M) \left( 1 + \frac{x_R - f_M}{x_R} \right)} \\
	& = \lim_{n \to \infty} \frac{x_M + p_1}{p_1 \sqrt{n} \left( 1 + \frac{p_1}{x_M + p_1} \right)} \\
	& = \lim_{n \to \infty} \frac{\frac{n+1}{2} + p_1}{p_1 \sqrt{n} \left( 1 + \frac{p_1}{\frac{n+1}{2} + p_1} \right)} \\
	& = \lim_{n \to \infty} \frac{n + 1 + 2 p_1}{2 p_1 \sqrt{n} \left( 1 + \frac{2 p_1}{n + 1 + 2 p_1} \right)} \\
	& = \lim_{n \to \infty} \frac{n + 1 + 2 p_1}{2 p_1 \sqrt{n}} \\
	& = \lim_{n \to \infty} \frac{\sqrt{n} + \frac{1}{\sqrt{n}} + 2 \frac{p_1}{\sqrt{n}}}{2 p_1} \\
	& = \lim_{n \to \infty} \frac{2.195 + \sqrt{n}}{2 p_1} \\
	& = \lim_{n \to \infty} \frac{\sqrt{n}}{2 p_1} = \frac{1}{2(1.0975)} = \frac{1}{2.195} \approx 0.456 \: .
\end{split}
\end{equation}

Now continue simplifying Equation~\ref{eq:p5limit1}, utilizing Equations~\ref{eq:c5limit},~\ref{eq:p1p5limit}, 
and~\ref{eq:lambdap5limit2} as follows:
\begin{equation}\label{eq:p5limit2}
\begin{split}
\lim_{n \to \infty} E[V] & \approx \lim_{n \to \infty} 1.596 \left(\frac{p_1 (1+2c)}{\sqrt{n}} + \frac{2c}{\sqrt{n}\lambda_R} \right) \\
	& \approx \lim_{n \to \infty} 1.596 \left( 1.0975 (1+2c) + 0.456 c \right) \\
	& \approx 1.596 \left( 1.0975 \cdot 1.268 + 0.456 \cdot 0.134 \right) \\
	& \approx 2.319 .
\end{split}
\end{equation}

Thus, when $p=0.5$ the expected number of uniform random variates required by BTPE to generate
one binomial random variate converges to approximately 2.319 as $n$ grows large 
(approximately 1.16 rejection sampling iterations on average). Observe in Table~\ref{tab:predicted}, 
that this limit case is reached quickly.

\section{Experimental Validation}\label{sec:experiments}

In earlier work, I used the open source Java library $\rho\mu$ to empirically explore the 
behavior of my BTPE implementation~\cite{cicirello2023asec}.
I wrapped an instance of Java's \textsc{SplittableRandom} class, which implements the 
splitmix~\cite{Steele2014} pseudorandom number generator (PRNG). The purpose was to instrument
the PRNG in order to count the number of uniform random variates, $U(0,1)$, generated
(i.e., calls to the \textsc{nextDouble()} method) while generating a binomial random variate. 
I then supply this wrapped PRNG instance as the source of randomness for $\rho\mu$'s 
implementation of BTPE.

That prior paper~\cite{cicirello2023asec} considered many combinations of $n$ and $p$. Here
I focus on the same values of $n \in \{ 2^{5}, 2^{6}, \ldots, 2^{20} \}$. But, I only focus on
two specific cases of $p$, the two cases utilized earlier in the paper:
$p=\frac{10}{n}$ from Section~\ref{sec:maxiter} and $p=0.5$ from Section~\ref{sec:miniter}.
For each binomial distribution $B(n,p)$, I generate 10,000 binomial random variates,
and I compute the average number of uniform variates per binomial, with 95\% confidence 
intervals. I compare the experimental mean to the prediction of the number of uniform variates
from Equation~\ref{eq:uniform}, and test significance with a $t$-test.
The experiments used OpenJDK 17 on a Windows 10 PC with a 3.4 GHz AMD A10-5700 CPU and 8 GB RAM. 
I used $\rho\mu$ 3.1.1. Both the source code for the experiments 
(\url{https://github.com/cicirello/btpe-iterations}), as well as for 
$\rho\mu$ (\url{https://github.com/cicirello/rho-mu}) is maintained on GitHub.
The API documentation for the $\rho\mu$ library is available on the web:
\url{https://rho-mu.cicirello.org/}.

\begin{table}[t]
\centering
\caption{Average number of calls to $U(0,1)$ by $\rho\mu$'s BTPE implementation
compared to the prediction of Equation~\ref{eq:uniform}. 95\% confidence intervals
are shown, as well as the $p$-values from $t$-tests.}\label{fig:results}
\begin{tabular}{c|ccc|ccc}\hline
    & \multicolumn{3}{|c}{$B(n,\frac{10}{n})$} & \multicolumn{3}{|c}{$B(n,0.5)$} \\
$n$ & predicted & mean & $p$-value & predicted & mean & $p$-value \\\hline
$2^{5}$ & 3.837 & $3.829 \pm 0.052$ & 0.76 & 3.599 & $3.582 \pm 0.046$ & 0.46 \\
$2^{6}$ & 3.792 & $3.786 \pm 0.050$ & 0.80 & 3.337 & $3.334 \pm 0.042$ & 0.89 \\
$2^{7}$ & 3.790 & $3.810 \pm 0.051$ & 0.45 & 2.985 & $3.004 \pm 0.034$ & 0.27 \\
$2^{8}$ & 3.793 & $3.765 \pm 0.051$ & 0.27 & 2.632 & $2.637 \pm 0.025$ & 0.66 \\
$2^{9}$ & 3.796 & $3.796 \pm 0.051$ & 0.99 & 2.420 & $2.412 \pm 0.019$ & 0.41 \\
$2^{10}$ & 3.797 & $3.818 \pm 0.051$ & 0.42 & 2.317 & $2.304 \pm 0.016$ & 0.11 \\
$2^{11}$ & 3.798 & $3.875 \pm 0.052$ & 0.00 & 2.307 & $2.308 \pm 0.017$ & 0.85 \\
$2^{12}$ & 3.798 & $3.808 \pm 0.051$ & 0.72 & 2.276 & $2.276 \pm 0.015$ & 0.98 \\
$2^{13}$ & 3.799 & $3.809 \pm 0.051$ & 0.70 & 2.300 & $2.288 \pm 0.016$ & 0.13 \\
$2^{14}$ & 3.799 & $3.798 \pm 0.051$ & 0.97 & 2.299 & $2.304 \pm 0.017$ & 0.55 \\
$2^{15}$ & 3.799 & $3.808 \pm 0.052$ & 0.73 & 2.300 & $2.286 \pm 0.016$ & 0.08 \\
$2^{16}$ & 3.799 & $3.828 \pm 0.052$ & 0.27 & 2.302 & $2.306 \pm 0.017$ & 0.65 \\
$2^{17}$ & 3.799 & $3.825 \pm 0.052$ & 0.32 & 2.310 & $2.301 \pm 0.016$ & 0.29 \\
$2^{18}$ & 3.799 & $3.818 \pm 0.051$ & 0.46 & 2.310 & $2.313 \pm 0.017$ & 0.73 \\
$2^{19}$ & 3.799 & $3.792 \pm 0.050$ & 0.80 & 2.313 & $2.314 \pm 0.017$ & 0.91 \\
$2^{20}$ & 3.799 & $3.807 \pm 0.051$ & 0.74 & 2.314 & $2.309 \pm 0.017$ & 0.58 \\
\hline
\end{tabular}
\end{table}

Table~\ref{fig:results} shows the results. The raw and processed data are 
available on GitHub (\url{https://github.com/cicirello/btpe-iterations}). 
The empirical results confirm the analytical prediction of 
Equation~\eqref{eq:uniform}. Observe that there is no significant difference between the 
analytical prediction and the empirically computed means. T-test $p$-values are 
above 0.05 in almost all cases (well above in most cases). There is only
a single case, $B(2^{11}, 10/2^{11})$, in Table~\ref{fig:results} where a 
$t$-test $p$-value is less than 0.05, seemingly suggesting that the implementation
behaves differently than theory predicts in this case. However, this case is 
explainable by random chance. There are 32 cases represented in 
Table~\ref{fig:results}, so this one case is approximately 3\% of the cases tested.
My more in depth empirical study explored a much larger number of cases
experimentally, finding no significant difference between predicted and observed
behavior in all but 4\% of cases~\cite{cicirello2023asec}. At significance level 0.05,
we should expect false-positives in approximately 5\% of cases.

\section{Discussion and Conclusions}\label{sec:conclusions}

In this report, I analyzed the runtime behavior of the BTPE algorithm for binomial
random variate generation, and observed the following concerning average performance:
\begin{itemize}
\item The average number of acceptance--rejection sampling iterations, $E[I]$, and
average number of uniform random numbers generated while generating a binomial random
variate, $E[V]$, are as follows:
\begin{gather}
1 \leq E[I] < 2 ,\\
2 \leq E[V] < 4 .
\end{gather}
At least one iteration is necessary (the above lower bound), and the above upper bounds
derive from the case of minimum supported $n=20$ for $p=0.5$, which is the case requiring
the maximum number of rejection sampling iterations on average (Table~\ref{tab:predicted}).

\item For large $n$ and the case of $p=\frac{10}{n}$, which maximizes the
expected number of rejection sampling iterations, we find that the expected number
of uniform random numbers required to generate one binomial random variate is a constant:
\begin{equation}
\lim_{n \to \infty} E[V] \approx 3.801 .
\end{equation}

\item For large $n$ and the case of $p=0.5$, the case that minimizes the expected number
of rejection sampling iterations, we find that the expected number of uniform random 
numbers required to generate one binomial random variate is also a constant:
\begin{equation}
\lim_{n \to \infty} E[V] \approx 2.319 .
\end{equation}
\end{itemize}
Thus, the average runtime of BTPE is $\Theta(1)$, while the average runtime of
many of the alternative algorithms for generating binomial random variates is
worse than a constant. For example, the average runtime of the inverse transform
method is $O(np)$~\cite{Kachitvichyanukul1988,Knuth3}.

In this report, I also experimentally validated my limit analysis
of Kachitvichyanukul and Schmeiser's formula, using my implementation of
BTPE in the $\rho\mu$ Java library~\cite{cicirello2022joss}. This validates:
(a) Kachitvichyanukul and Schmeiser's formula for the expected number
of rejection sampling iterations, (b) my analysis of that formula, and
(c) that $\rho\mu$'s implementation of BTPE behaves as theory predicts. One
limitation of this study is that it focuses on average case performance. It is
possible that BTPE may exhibit many more rejection sampling iterations on
occasion. The maximum number of uniform random variates generated while
generating a single binomial during the course of my previous 
experimentation~\cite{cicirello2023asec} was 38 (19 rejection sampling iterations).
Such instances were very rare in that study, which generated approximately 2.88
million binomial random variates.

\bibliographystyle{plainnat}
\bibliography{binomial}

\begin{thebibliography}{22}
\providecommand{\natexlab}[1]{#1}
\providecommand{\url}[1]{\texttt{#1}}
\expandafter\ifx\csname urlstyle\endcsname\relax
  \providecommand{\doi}[1]{doi: #1}\else
  \providecommand{\doi}{doi: \begingroup \urlstyle{rm}\Url}\fi

\bibitem[Ahrens and Dieter(1974)]{Ahrens1974}
J.~H. Ahrens and U.~Dieter.
\newblock Computer methods for sampling from gamma, beta, poisson and bionomial
  distributions.
\newblock \emph{Computing}, 12\penalty0 (3):\penalty0 223--246, 1974.
\newblock \doi{10.1007/BF02293108}.

\bibitem[Arshad et~al.(2019)Arshad, Khan, Sharif, Yasmin, and
  Javed]{Arshad2019}
Habiba Arshad, Muhammad~Attique Khan, Muhammad Sharif, Mussarat Yasmin, and
  Muhammad~Younus Javed.
\newblock Multi-level features fusion and selection for human gait recognition:
  an optimized framework of bayesian model and binomial distribution.
\newblock \emph{International Journal of Machine Learning and Cybernetics},
  10\penalty0 (12):\penalty0 3601--3618, 2019.
\newblock \doi{10.1007/s13042-019-00947-0}.

\bibitem[Cicirello(2018)]{cicirello2018joss}
Vincent~A. Cicirello.
\newblock {JavaPermutationTools}: A java library of permutation distance
  metrics.
\newblock \emph{Journal of Open Source Software}, 3\penalty0 (31):\penalty0
  950, November 2018.
\newblock \doi{10.21105/joss.00950}.

\bibitem[Cicirello(2020)]{cicirello2020joss}
Vincent~A. Cicirello.
\newblock Chips-n-salsa: A java library of customizable, hybridizable,
  iterative, parallel, stochastic, and self-adaptive local search algorithms.
\newblock \emph{Journal of Open Source Software}, 5\penalty0 (52):\penalty0
  2448, August 2020.
\newblock \doi{10.21105/joss.02448}.

\bibitem[Cicirello(2022)]{cicirello2022joss}
Vincent~A. Cicirello.
\newblock $\rho\mu$: A java library of randomization enhancements and other
  math utilities.
\newblock \emph{Journal of Open Source Software}, 7\penalty0 (76):\penalty0
  4663, August 2022.
\newblock \doi{10.21105/joss.04663}.

\bibitem[Cicirello(2023)]{cicirello2023asec}
Vincent~A. Cicirello.
\newblock An analysis of an open source binomial random variate generation
  algorithm.
\newblock \emph{Engineering Proceedings}, 56\penalty0 (1):\penalty0 86, October
  2023.
\newblock \doi{10.3390/ASEC2023-15349}.

\bibitem[Cormen et~al.(2022)Cormen, Leiserson, Rivest, and Stein]{CLRS}
Thomas~H. Cormen, Charles~E. Leiserson, Ronald~L. Rivest, and Clifford Stein.
\newblock \emph{Introduction to Algorithms}.
\newblock MIT Press, Cambridge, MA, 4th edition, 2022.

\bibitem[Flury(1990)]{Flury1990}
Bernard~D. Flury.
\newblock Acceptance–rejection sampling made easy.
\newblock \emph{SIAM Review}, 32\penalty0 (3):\penalty0 474--476, 1990.
\newblock \doi{10.1137/1032082}.

\bibitem[García-García et~al.(2022)García-García, Fernández~Coronado,
  Arredondo, and Imilpán~Rivera]{Garcia2022}
Jaime~Israel García-García, Nicolás~Alonso Fernández~Coronado, Elizabeth~H.
  Arredondo, and Isaac~Alejandro Imilpán~Rivera.
\newblock The binomial distribution: Historical origin and evolution of its
  problem situations.
\newblock \emph{Mathematics}, 10\penalty0 (15):\penalty0 2680, 2022.
\newblock \doi{10.3390/math10152680}.

\bibitem[Kachitvichyanukul and Schmeiser(1988)]{Kachitvichyanukul1988}
Voratas Kachitvichyanukul and Bruce~W. Schmeiser.
\newblock Binomial random variate generation.
\newblock \emph{Communications of the ACM}, 31\penalty0 (2):\penalty0 216--222,
  1988.
\newblock \doi{10.1145/42372.42381}.

\bibitem[Kachitvichyanukul and Schmeiser(1989)]{Kachitvichyanukul1989}
Voratas Kachitvichyanukul and Bruce~W. Schmeiser.
\newblock Algorithm 678: Btpec: Sampling from the binomial distribution.
\newblock \emph{ACM Transactions on Mathematical Software}, 15\penalty0
  (4):\penalty0 394--397, 1989.
\newblock \doi{10.1145/76909.76916}.

\bibitem[Khan and Olivier(2019)]{Khan2019}
Manzoor Khan and Jake Olivier.
\newblock Regression to the mean for the bivariate binomial distribution.
\newblock \emph{Statistics in Medicine}, 38\penalty0 (13):\penalty0 2391--2412,
  2019.
\newblock \doi{10.1002/sim.8115}.

\bibitem[Knuth(1998)]{Knuth3}
Donald~E. Knuth.
\newblock \emph{The Art of Computer Programming, Volume 2, Seminumerical
  Algorithms}.
\newblock Addison Wesley, 3rd edition, 1998.

\bibitem[Kuhl(2017)]{Kuhl2017}
Michael~E. Kuhl.
\newblock History of random variate generation.
\newblock In \emph{Proceedings of the 2017 Winter Simulation Conference}, pages
  231--242. IEEE Press, 2017.
\newblock \doi{10.1109/WSC.2017.8247791}.

\bibitem[Naimi and Whitcomb(2020)]{Naimi2020}
Ashley~I Naimi and Brian~W Whitcomb.
\newblock Estimating risk ratios and risk differences using regression.
\newblock \emph{American Journal of Epidemiology}, 189\penalty0 (6):\penalty0
  508--510, 2020.
\newblock \doi{10.1093/aje/kwaa044}.

\bibitem[Relles(1972)]{Relles1972}
Daniel~A. Relles.
\newblock A simple algorithm for generating binomial random variables when n is
  large.
\newblock \emph{Journal of the American Statistical Association}, 67\penalty0
  (339):\penalty0 612--613, 1972.
\newblock \doi{10.1080/01621459.1972.10481259}.

\bibitem[Shah et~al.(2022)Shah, Khan, and Ahmad]{Shah2022}
Sayed Qaiser~Ali Shah, Farrukh~Zeeshan Khan, and Muneer Ahmad.
\newblock Mitigating tcp syn flooding based edos attack in cloud computing
  environment using binomial distribution in sdn.
\newblock \emph{Computer Communications}, 182:\penalty0 198--211, 2022.
\newblock \doi{10.1016/j.comcom.2021.11.008}.

\bibitem[Singh et~al.(2022)Singh, Chawla, Prasad, Anand, Alharbi, and
  Alosaimi]{Singh2022}
Sunny Singh, Muskaan Chawla, Devendra Prasad, Divya Anand, Abdullah Alharbi,
  and Wael Alosaimi.
\newblock An improved binomial distribution-based trust management algorithm
  for remote patient monitoring in wbans.
\newblock \emph{Sustainability}, 14\penalty0 (4):\penalty0 2141, 2022.
\newblock \doi{10.3390/su14042141}.

\bibitem[Steele(2017)]{Steele2017}
Guy Steele.
\newblock Jep 356: Enhanced pseudo-random number generators.
\newblock JEP 356, OpenJDK, 2017.
\newblock URL \url{https://openjdk.org/jeps/356}.

\bibitem[Steele et~al.(2014)Steele, Lea, and Flood]{Steele2014}
Guy~L. Steele, Doug Lea, and Christine~H. Flood.
\newblock Fast splittable pseudorandom number generators.
\newblock In \emph{Proceedings of the 2014 ACM International Conference on
  Object Oriented Programming Systems Languages \& Applications}, pages
  453--472, New York, NY, USA, 2014. Association for Computing Machinery.
\newblock \doi{10.1145/2660193.2660195}.

\bibitem[Wang and Pei(2019)]{Wang2019}
Guantao Wang and Jingjing Pei.
\newblock Macro risk: A versatile and universal strategy for measuring the
  overall safety of hazardous industrial installations in china.
\newblock \emph{International Journal of Environmental Research and Public
  Health}, 16\penalty0 (10):\penalty0 1680, 2019.
\newblock \doi{10.3390/ijerph16101680}.

\bibitem[Zhang and Lin(2020)]{Zhang2020}
Xiaoxian Zhang and Zhifen Lin.
\newblock Hormesis-induced gap between the guidelines and reality in ecological
  risk assessment.
\newblock \emph{Chemosphere}, 243:\penalty0 125348, 2020.
\newblock \doi{10.1016/j.chemosphere.2019.125348}.

\end{thebibliography}
\end{document}